\documentclass[english,10pt,aps,prd,a4paper,preprintnumbers,floatfix,nofootinbib,showpacs,superscriptaddress, twocolumn]{revtex4-1}
\usepackage[english]{babel}
\usepackage{amsmath}
\usepackage{graphicx}
\usepackage{color}
\usepackage{mathrsfs}   
\usepackage{amssymb}
\usepackage{hyperref}
\usepackage[normalem]{ulem} 
\usepackage{dcolumn}
\usepackage{bm}
\usepackage{graphicx}
\usepackage[sort&compress]{natbib}

\definecolor{greenLinks}{rgb}{0, 0.6, 0}
\definecolor{blueLinks}{rgb}{0, 0, 0.6}
\definecolor{redLinks}{rgb}{0.6, 0, 0}
\definecolor{tempText}{rgb}{0.55, 0.10,0.67}
\definecolor{eprintLinks}{rgb}{0.4, 0.4, 0.4}
\definecolor{journalLinks}{rgb}{0.6, 0, 0}

\usepackage{float}

\newcommand{\AddrAHEP}{%
  AHEP Group, Institut de F\'{i}sica Corpuscular --
  C.S.I.C./Universitat de Val\`{e}ncia, Parc Cient\'ific de Paterna.\\
 C/ Catedr\'atico Jos\'e Beltr\'an, 2 E-46980 Paterna (Valencia) - SPAIN}

\begin{document}

\title{Realistic Tri-Bi-Maximal neutrino mixing}

\author{Peng Chen}\email{pche@mail.ustc.edu.cn}
\affiliation{College of Information Science and Engineering,Ocean University of China, Qingdao 266100, China}
\author{Salvador Centelles Chuli\'{a}}\email{salcen@ific.uv.es}
\affiliation{\AddrAHEP}
\author{Gui-Jun Ding}\email{dinggj@ustc.edu.cn}
\affiliation{Interdisciplinary Center for Theoretical Study and Department of Modern Physics, \\
University of Science and Technology of China, Hefei, Anhui 230026, China}
\author{Rahul Srivastava}\email{rahulsri@ific.uv.es}
\affiliation{\AddrAHEP}
\author{Jos\'{e} W. F. Valle}\email{valle@ific.uv.es}
\affiliation{\AddrAHEP}

\begin{abstract}
  \vspace{1cm}

We propose a generalized version of the Tri-Bi-Maximal (TBM) ansatz for lepton mixing, leading to non-zero reactor angle $\theta_{13}$ and CP violation.
The latter is characterized by two CP phases. The Dirac phase affecting neutrino oscillations is nearly maximal ($\delta_{CP} \sim \pm \pi/2$), while the Majorana phase implies narrow allowed ranges for the neutrinoless double beta decay amplitude.
The solar angle $\theta_{12}$ lies nearly at its TBM value, while the atmospheric angle $\theta_{23}$ has the TBM value for maximal $\delta_{CP}$.
Neutrino oscillation predictions can be tested in present and upcoming experiments.

\end{abstract}

\preprint{USTC-ICTS-18-11}

\maketitle


Ever since the discovery of neutrino oscillations, the structure of leptonic mixing matrix has been an active topic of research.
Over the last twenty years or so, there has been a flood of both theoretical and experimental activity aimed at determining and understanding the structure of leptonic mixing matrix.
Solar and atmospheric data, confirmed by accelerator and reactor data made it clear that the structure of lepton mixing is quite at odds with that of quarks, given the large values of $\theta_{12}$ and $\theta_{23}$. These observations were soon encoded in the Tri-Bi-Maximal Mixing (TBM) ansatz proposed by Harrison, Perkins, and Scott~\cite{Harrison:2002er}, described by\\[-.3cm]
 \begin{eqnarray}
 U_{0} & = &  \left[
\begin{array}{ccc}
\sqrt{\frac{2}{3}}               & \frac{1}{\sqrt{3}}           & 0   \\
-\frac{1}{\sqrt{6}}              & \frac{1}{\sqrt{3}}           & \frac{1}{\sqrt{2}} \\
\frac{1}{\sqrt{6}}               &  -\frac{1}{\sqrt{3}}         & \frac{1}{\sqrt{2}} \\
\end{array}
\right]
\label{eq:tbm}
\end{eqnarray}

\vskip .1cm
Since it was first proposed, the TBM ansatz has been a popular benchmark for describing the pattern of lepton mixing, inspiring a flood of theory papers.
It gives $ \theta_{12} = \sin^{-1} \left(\frac{1}{\sqrt{3}}\right)$,
$\theta_{23} = \pi/4$ whose status is rather good in view of the latest neutrino oscillation global fit~\cite{deSalas:2017kay, globalfit}.
Unfortunately, it predicts $\theta_{13}=0$ and hence CP-conservation in neutrino oscillation.
Indeed, data from reactors have indicated that such ``bona-fide'' TBM ansatz can not be the correct description of nature, since the leptonic mixing angle $\theta_{13}$ has been established to be non-zero to a very high significance~\cite{An:2016ses,Pac:2018scx,Abe:2014bwa}.
Moreover, there has been mounting evidence for CP violation in neutrino oscillations, providing further indication that amendment is needed. \\[-.2cm]

Motivated by the need for departing from the simplest ``first-order'' form for the TBM ansatz, Eq.~\ref{eq:tbm}, here we propose a generalized version of the TBM ansatz (gTBM) which correctly accounts for the non-zero value of $\theta_{13}$ and introduces CP violation as follows \\[-.3cm]
\begin{eqnarray}
\label{eq:utbsym}
U & = &  \left[
\begin{array}{ccc} \sqrt{\frac{2}{3}}    & \frac{ e^{-i \rho } \cos \theta}{\sqrt{3}}     & -\frac{i e^{-i \rho } \sin \theta}{\sqrt{3}}
\\
-\frac{e^{i \rho }}{\sqrt{6}}                          & \frac{\cos \theta}{\sqrt{3}} - \frac{i e^{-i\sigma} \sin \theta}{\sqrt{2}}
& \frac{e^{-i\sigma} \cos \theta}{\sqrt{2}} - \frac{i \sin \theta}{\sqrt{3}}
\\
\frac{e^{i (\rho + \sigma)}}{\sqrt{6}}         &  -\frac{e^{i\sigma} \cos \theta}{\sqrt{3}} - \frac{i \sin \theta}{\sqrt{2}}
& \frac{\cos \theta}{\sqrt{2}} + \frac{i e^{i \sigma} \sin \theta}{\sqrt{3}}
\\
\end{array}
\right]
\end{eqnarray} \\[-.2cm]

This new ansatz is characterized by just one angle $\theta$ and two phases $\rho$ and $\sigma$. These are three parameters, to be compared with the three angles plus three (physical) phases characterizing the three-family (unitary) lepton mixing matrix~\cite{Schechter:1980gr}.
The latter can be written in symmetric form as $ U = U_{23}(\theta_{23},~\phi_{23}) \cdot U_{13} (\theta_{13},~\phi_{13}) \cdot U_{12}(\theta_{12},~\phi_{12})$,
where $U_{ij}(\theta, \phi)$ are matrices corresponding to complex rotations in the $ij$ plane, each characterized by an angle $\theta_{ij}$ and an associated phase $\phi_{ij}$ ~\cite{Schechter:1980gr}. In addition to the Dirac CP phase $\delta_{CP}  =  \phi_{13} - \phi_{12} - \phi_{23}$  one has two Majorana phases~\cite{Schechter:1980gk,Doi:1980yb} that affect neutrinoless double beta decay.
Eq.~\ref{eq:utbsym} gives all of these six parameters in terms of one angle $\theta$ plus two phase parameters $\rho, \sigma$. The parameters have ranges
\begin{eqnarray}
 0 \leq \theta < \pi, \, 0 \leq \rho < \pi, \, 0 \leq \sigma < 2 \pi
 \label{phyrange}
\end{eqnarray}
We now turn to the several interesting limiting cases of the above gTBM matrix in Eq.~\eqref{eq:utbsym}.


\subsection{TBM Limit} The first is the limit $\theta, \rho, \sigma \to 0$, in which case our gTBM mixing matrix in Eq.~\eqref{eq:utbsym} reduces to the simplest celebrated TBM form, $U_0$ in Eq.~\eqref{eq:tbm}. This is unrealistic, as it can not describe reactor neutrino data.


\subsection{Complex TBM Limit}  In the limit of $\theta \to 0$ and any arbitrary value of $\rho, \sigma $, the matrix reduces to  ``complex TBM'' matrix which is TBM matrix with additional CP phases. This matrix is given by
\begin{eqnarray}
 U & = &  \left[
\begin{array}{ccc}
\sqrt{\frac{2}{3}}                       & \frac{ e^{-i \rho } }{\sqrt{3}}     & 0                                \\
-\frac{e^{i \rho }}{\sqrt{6}}            & \frac{1}{\sqrt{3}}                  & \frac{e^{-i\sigma}}{\sqrt{2}}    \\
\frac{e^{i (\rho + \sigma)}}{\sqrt{6}}   &  -\frac{e^{i\sigma}}{\sqrt{3}}      & \frac{1}{\sqrt{2}}               \\
\end{array}
\right]
\label{eq:ctbm}
\end{eqnarray}
The phases $\rho$ and $\sigma$ are physical parameters only if neutrinos are Majorana-type, and can be rotated away otherwise. Indeed, for Dirac neutrinos there is no difference between TBM and complex TBM. For the Majorana neutrino case the phases in the symmetric parametrization are given as $\phi_{12}  = \rho$ and $\phi_{23}  = \sigma$, while the Dirac phase $\delta_{CP}$ is unphysical, since $\theta_{13} =0$.


\subsection{The $\mu - \tau$ Symmetric Limit}

We now discuss realistic limits of gTBM that lead to $\theta_{13} \neq 0$, as required by current data~\cite{An:2016ses,Pac:2018scx,Abe:2014bwa}.
One of the properties of the TBM matrix was the so-called $\mu - \tau$ symmetry, i.e. $|U_{\mu j}| = |U_{\tau j}|$; $j = 1,2,3$~\cite{Harrison:2002er,Grimus:2003yn}.
For $\sigma \to 0$ and any arbitrary values of $\theta, \rho$, the gTBM matrix also retains this symmetry, reducing to
\begin{eqnarray}
 U & = &
 \left[
\begin{array}{ccc}
\sqrt{\frac{2}{3}}                                     & \frac{ e^{-i \rho } \cos \theta}{\sqrt{3}}                                        & -\frac{i e^{-i \rho } \sin \theta}{\sqrt{3}}
\\
-\frac{e^{i \rho }}{\sqrt{6}}                          & \frac{\cos \theta}{\sqrt{3}} - \frac{i \sin \theta}{\sqrt{2}}
& \frac{\cos \theta}{\sqrt{2}} - \frac{i \sin \theta}{\sqrt{3}}
\\
\frac{e^{i \rho }}{\sqrt{6}}         &  -\frac{\cos \theta}{\sqrt{3}} - \frac{i \sin \theta}{\sqrt{2}}
& \frac{\cos \theta}{\sqrt{2}} + \frac{i \sin \theta}{\sqrt{3}}
\\
\end{array}
\right]
\label{eq:mu-tau}
\end{eqnarray}
Indeed, one sees that the matrix in \eqref{eq:mu-tau} also has an inherent $\mu-\tau$ symmetry, leading to maximal atmospheric angle $\theta_{23} = \frac{\pi}{4}$
and maximal CP violating value of CP phase $\delta_{CP} = \pm \frac{\pi}{2}$.
The other two angles are also non-zero and are correlated with each other, as follows
\begin{eqnarray}
\cos^2 \theta_{12} \cos^2 \theta_{13} = \frac{2}{3}\,.
\label{cor-the12-the13}
\end{eqnarray}

\begin{figure}[H]
\centering
\includegraphics[scale=0.6]{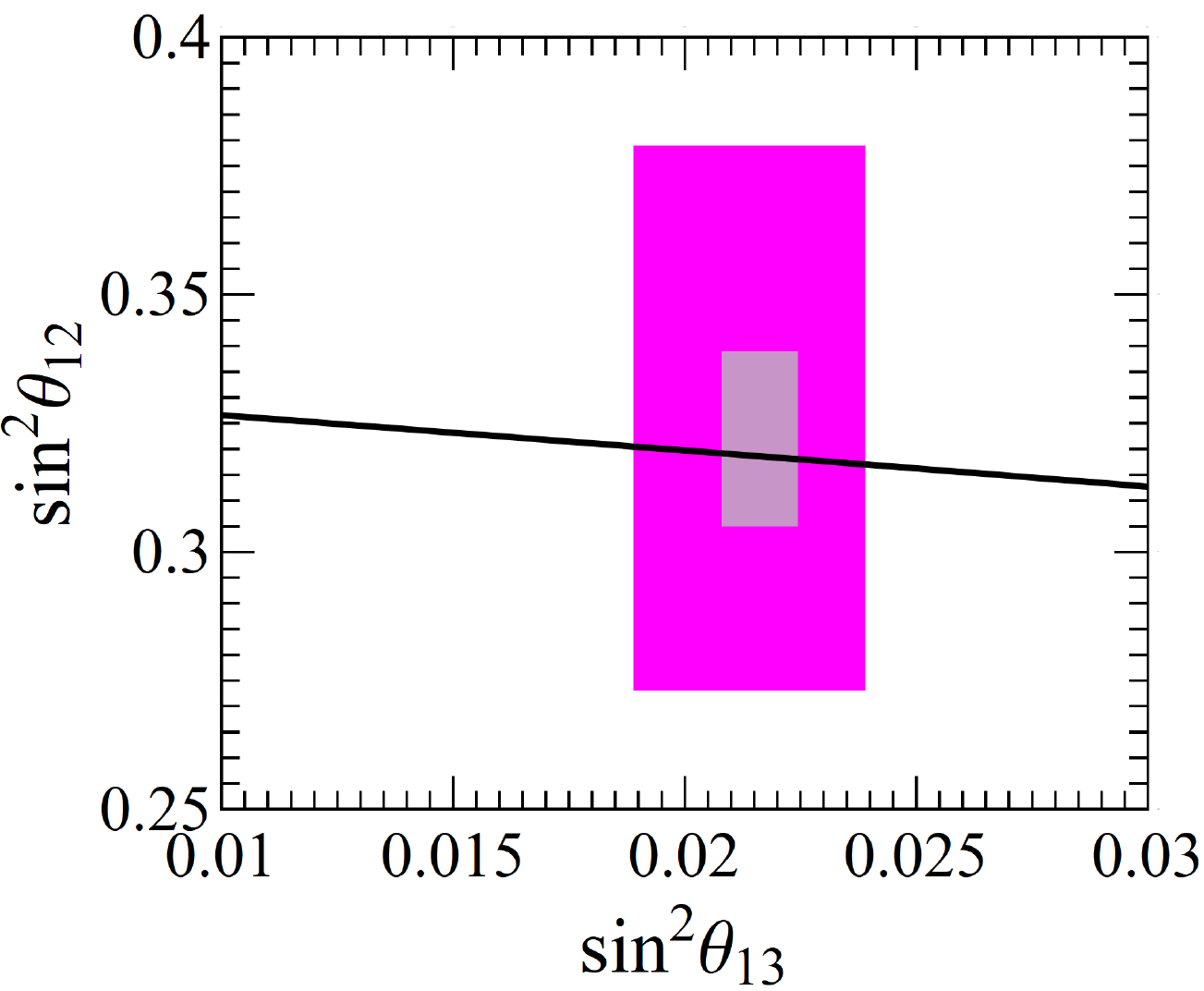}
\caption{Correlation between $\sin^2 \theta_{13}$  and $\sin^2 \theta_{12}$ given in Eq.~\ref{cor-the12-the13}. 
Notice that in the whole experimentally allowed range~\cite{deSalas:2017kay}, the value of $\sin^2 \theta_{12}$ remains very close to $1/3$.}
\label{fig:the12-the13}
\end{figure}

Using the $3\sigma$ range of the reactor mixing angle~$1.96\times10^{-2}\leq\sin^2\theta_{13}\leq2.41\times10^{-2}$~\cite{deSalas:2017kay, globalfit}, we obtain for the solar mixing angle $0.346\leq\sin^2\theta_{12}\leq0.349$. This is illustrated in Fig. \ref{fig:the12-the13}, in which the shaded boxes highlight the 1 and 3$\sigma$ regions indicated by the current neutrino oscillation global fit. This correlation is rather different from the one predicted in \cite{Morisi:2013qna}. The additional CP phases are physical, both Majorana and Dirac, since $\theta_{13} \neq 0$ makes $\phi_{13}$ also well defined.
This $\mu-\tau$ symmetric case has implications for $m_{ee}$, shown in the Fig. \ref{fig:mee-fig}.

In the $\mu-\tau$ symmetric matrix of Eq.~\eqref{eq:mu-tau}, one can further take the $\rho \to 0$ limit, in which case we get an even simpler matrix given by
\begin{eqnarray}
 U & = &
 \left[
\begin{array}{ccc}
\sqrt{\frac{2}{3}}                                     & \frac{ \cos \theta}{\sqrt{3}}                                        & -\frac{i  \sin \theta}{\sqrt{3}}
\\
-\frac{1}{\sqrt{6}}                          & \frac{\cos \theta}{\sqrt{3}} - \frac{i \sin \theta}{\sqrt{2}}
& \frac{\cos \theta}{\sqrt{2}} - \frac{i \sin \theta}{\sqrt{3}}
\\
\frac{1}{\sqrt{6}}         &  -\frac{\cos \theta}{\sqrt{3}} - \frac{i \sin \theta}{\sqrt{2}}
& \frac{\cos \theta}{\sqrt{2}} + \frac{i \sin \theta}{\sqrt{3}}
\\
\end{array}
\right]
\label{eq:real-mu-tau}
\end{eqnarray}

\begin{figure}[H]
 \centering
 \includegraphics[scale=0.45]{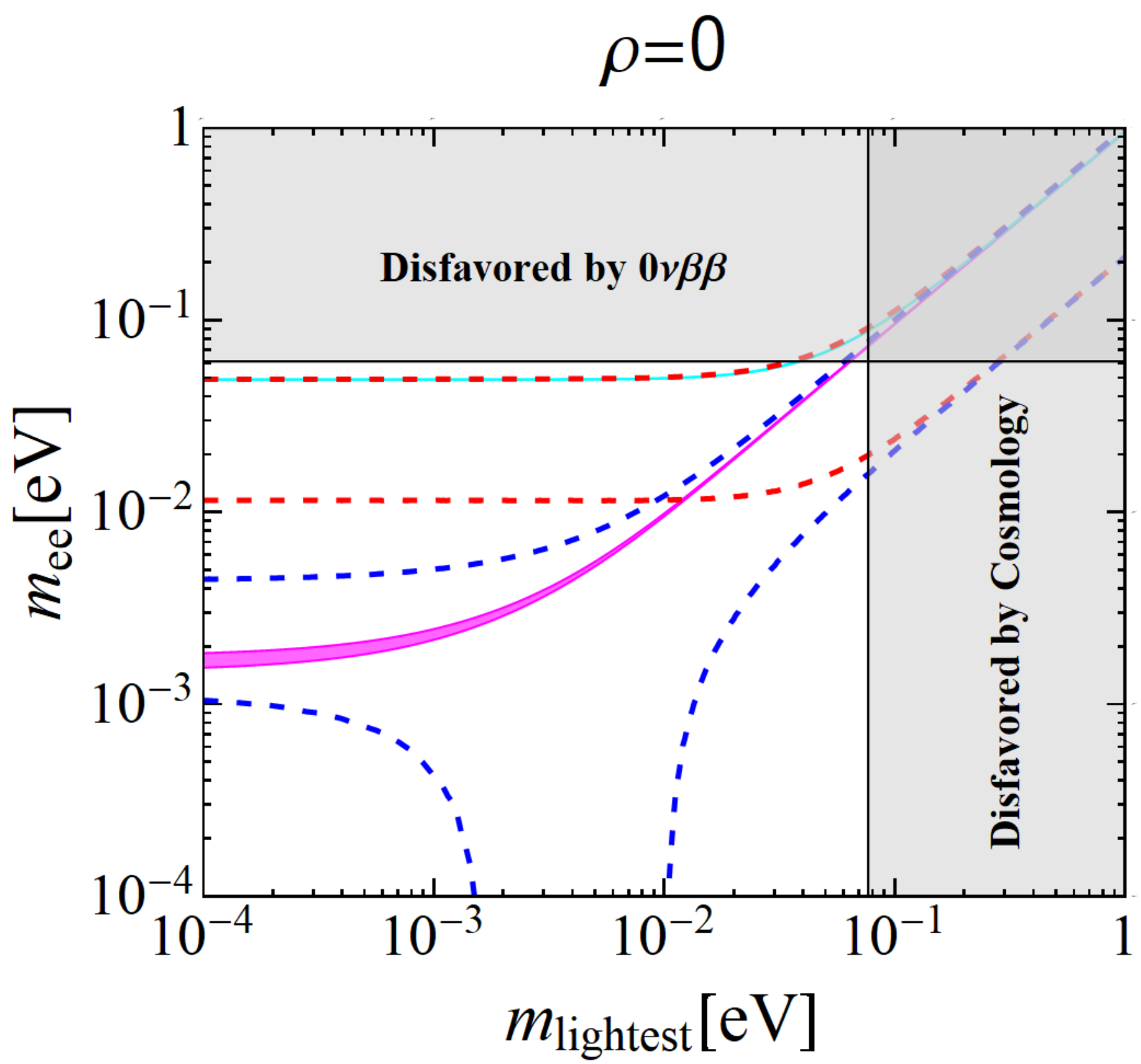}~~~~~~
\caption{$|m_{ee}|$ prediction for NO and IO when $\rho=0$. Here $\theta$ is taken as a free parameter, and we require the three mixing angles to lie in their allowed $3\sigma$ regions~\cite{deSalas:2017kay, globalfit}. Note that $m_{ee}$ does not depend on $\sigma$.} 
\label{fig:mee-rho0-fig}
\end{figure}

Notice that this matrix shares many properties of matrix in Eq.~\eqref{eq:mu-tau} e.g. maximal atmospheric angle, maximal CP violation and the correlation given in Eq.~\eqref{cor-the12-the13}.
In addition, the Majorana phase is fixed, since now $\rho =0$, leading to very sharp predictions for $m_{ee}$ as shown in Fig. \ref{fig:mee-rho0-fig}. For example, for the case of inverse ordering (IO) the neutrinoless double beta decay amplitude is nearly maximal, while for the NO case there is a lower bound for this amplitude, since destructive interference is prevented.

\subsection{The $\rho \to 0$ Limit}

\begin{figure}[H]
\centering
\includegraphics[scale=0.6]{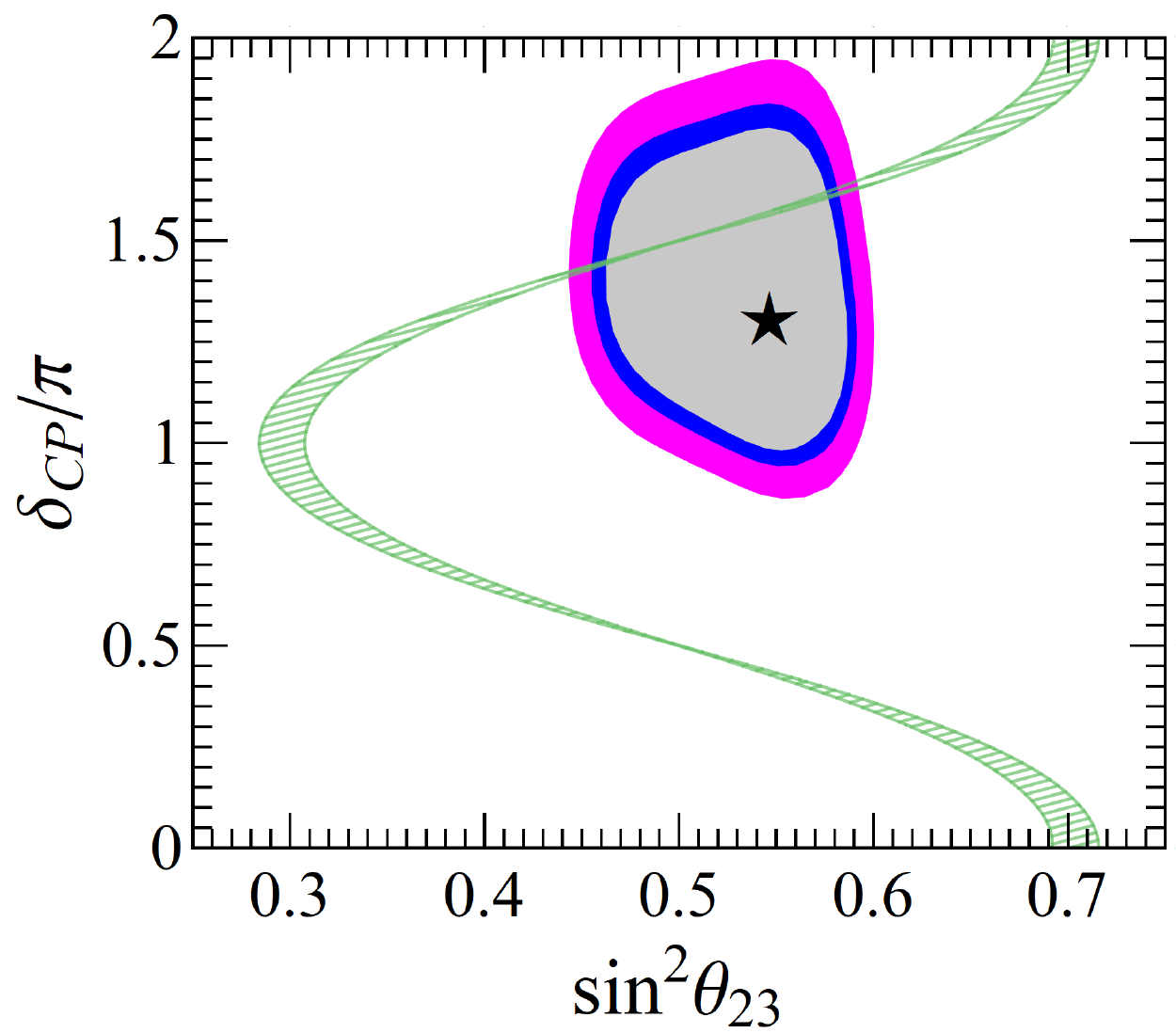}~~~~~~
\caption{The correlation between atmospheric angle $\theta_{23}$ and CP phase $\delta_{CP}$ predicted by our generalized TBM matrix in Eq.~\ref{eq:utbsym} is given by the hatched band, while the 1, 2 and 3$\sigma$ regions allowed by the current neutrino oscillation global fit are indicated by the shaded areas~\cite{deSalas:2017kay, globalfit}.}
\label{fig:the23-delta}
\end{figure}

 So far the limits we have discussed all lead to maximal atmospheric mixing angle i.e. they all predict $\theta_{23} = \pi/4$. While this is consistent with current data, there is a slight preference for the second octant~\cite{deSalas:2017kay, globalfit}. Our proposed gTBM matrix is flexible enough to allow for deviations from maximal $\theta_{23}$.
 The possibility of non-maximal $\theta_{23}$ can be seen in the limiting case where $\rho \to 0$, where the mixing matrix is given by
\begin{eqnarray}
\hskip-0.1in U & = &  \left[
\begin{array}{ccc} \sqrt{\frac{2}{3}}    & \frac{  \cos \theta}{\sqrt{3}}     & -\frac{i  \sin \theta}{\sqrt{3}}
\\
-\frac{1}{\sqrt{6}}                          & \frac{\cos \theta}{\sqrt{3}} - \frac{i e^{-i\sigma} \sin \theta}{\sqrt{2}}
& \frac{e^{-i\sigma} \cos \theta}{\sqrt{2}} - \frac{i \sin \theta}{\sqrt{3}}
\\
\frac{e^{i  \sigma}}{\sqrt{6}}         &  -\frac{e^{i\sigma} \cos \theta}{\sqrt{3}} - \frac{i \sin \theta}{\sqrt{2}}
& \frac{\cos \theta}{\sqrt{2}} + \frac{i e^{i \sigma} \sin \theta}{\sqrt{3}}
\\
\end{array}
\right]
\label{eq:rho0}
\end{eqnarray}
This matrix still shares some of the properties of the $\mu-\tau$ symmetric matrix of \eqref{eq:mu-tau}.
For example, the correlation in~\eqref{cor-the12-the13} still holds, relating solar and reactor angles as shown in Fig. \ref{fig:the12-the13}.
However, in contrast to the $\mu -\tau$ symmetric limit, we can now have deviations from maximal atmospheric mixing, as well as deviations from
maximal CP violation.
In fact, these departures are correlated with each other, as shown in Fig. \ref{fig:the23-delta}, which also highlights the 1 and 3$\sigma$ regions indicated by the current neutrino oscillation global fit~\cite{deSalas:2017kay, globalfit}.

The mixing matrix of \eqref{eq:rho0} also leads to fixed Majorana phase values given by $ \phi_{12}  =  0, \phi_{13} = \frac{\pi}{2}$ implying sharp predictions for $m_{ee}$, as shown in Fig. \ref{fig:mee-rho0-fig}.

\subsection{General Tri-Bi-Maximal Mixing}

\begin{figure}[H]
\centering
\includegraphics[scale=0.6]{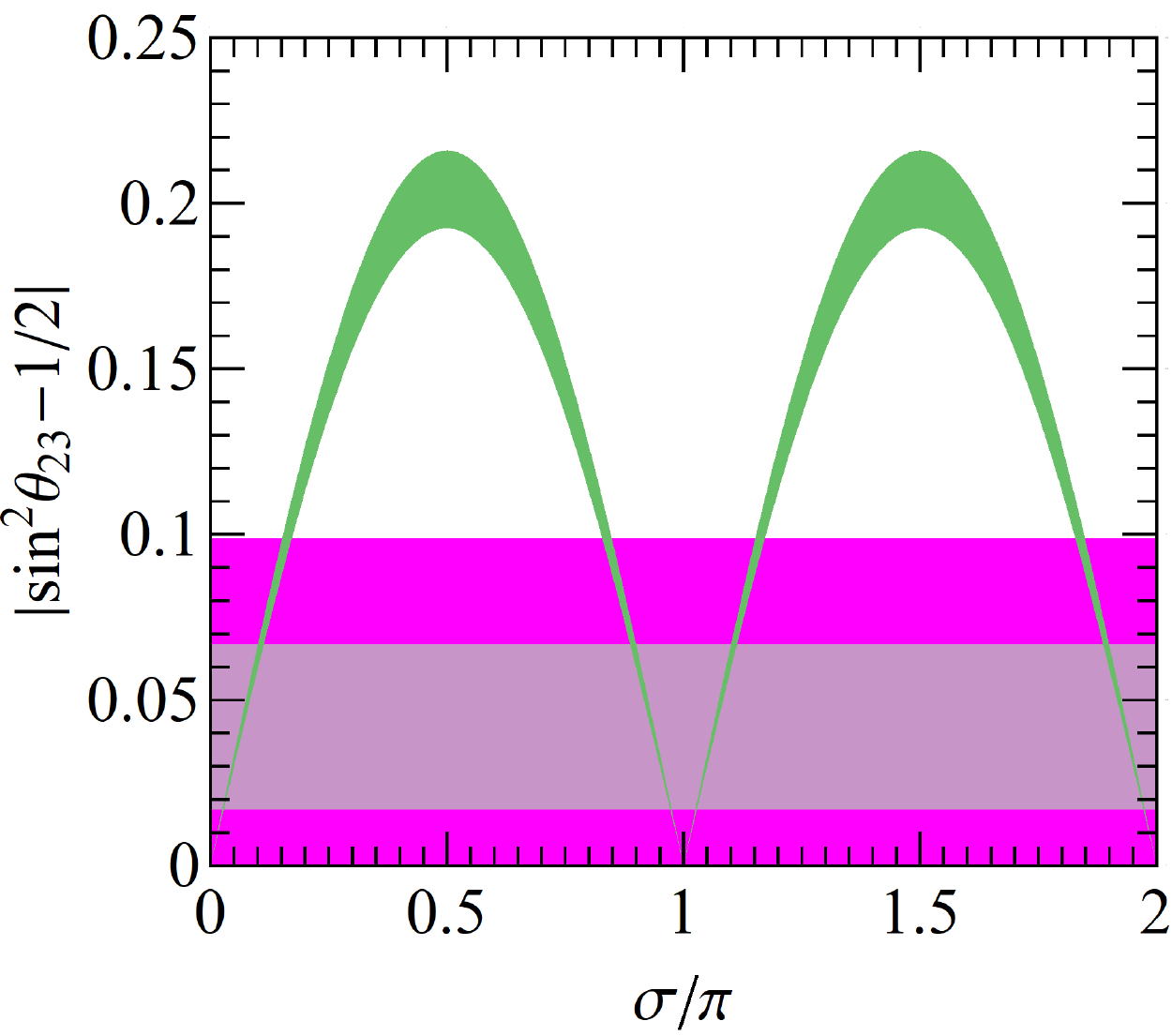}~~~~~~
\caption{The predicted dependence of $|\sin^2\theta_{23}-\frac{1}{2}|$ on the parameter $\sigma$ is indicated by the curved band. Its width comes from varying $\theta_{13}$ within its $3\sigma$ range, while the horizontal band gives the current determination of $\theta_{23}$~\cite{deSalas:2017kay, globalfit}. } 
\label{fig:theta23-sigma-fig}
\end{figure}

Having discussed the various limits of our proposed gTBM matrix, \eqref{eq:utbsym}, we now briefly discuss its general properties.
The full set of mixing angles and phases is given as
\begin{eqnarray}
& & \sin^2\theta_{12}  =  \frac{\cos^2\theta}{\cos^2\theta+2}\,,
\, \,
\sin^2\theta_{23} =  \frac{1}{2} + \frac{\sqrt{6}\sin  2\theta\sin\sigma}{2\cos^2\theta+4}\,,
\nonumber \\
& & \sin^2\theta_{13}  =  \frac{\sin^2\theta}{3}\,,
\, \,
\tan\delta_{CP} = \frac{(\cos^2\theta+2)\cot\sigma}{(5\cos^2\theta-2)}\,,
\nonumber \\
& & \phi_{12}    =  \rho \,,
\, \, \phi_{13} = \rho+\frac{\pi}{2}\,, \\ 
\label{eq:mix23}
\end{eqnarray}
which implies
\begin{equation}
|\sin^2\theta_{23}-\frac{1}{2}|=\tan\theta_{13}\sqrt{2-4\tan^2\theta_{13}}\,|\sin\sigma|\,.
\end{equation}
The parameter $\sigma$ measures the deviation of $\theta_{23}$ from maximal mixing, as shown in Fig.~\ref{fig:theta23-sigma-fig}. We can read off that $\sigma$ can only vary within the region $[0, 0.172\pi]\cup[0.828\pi, 1.172\pi]\cup[1.828\pi, 2\pi)$.

The expression for the parameter $m_{ee}$ describing the neutrinoless double beta decay amplitude also takes a rather simple form given by
\begin{eqnarray}
  |m_{ee}| & = &  \frac{1}{3} |2 e^{2 i \rho} m_1 + m_2 \cos^2 \theta - m_3 \sin^2 \theta|
\end{eqnarray}

\begin{figure}[h]
 \centering
 \includegraphics[scale=0.45]{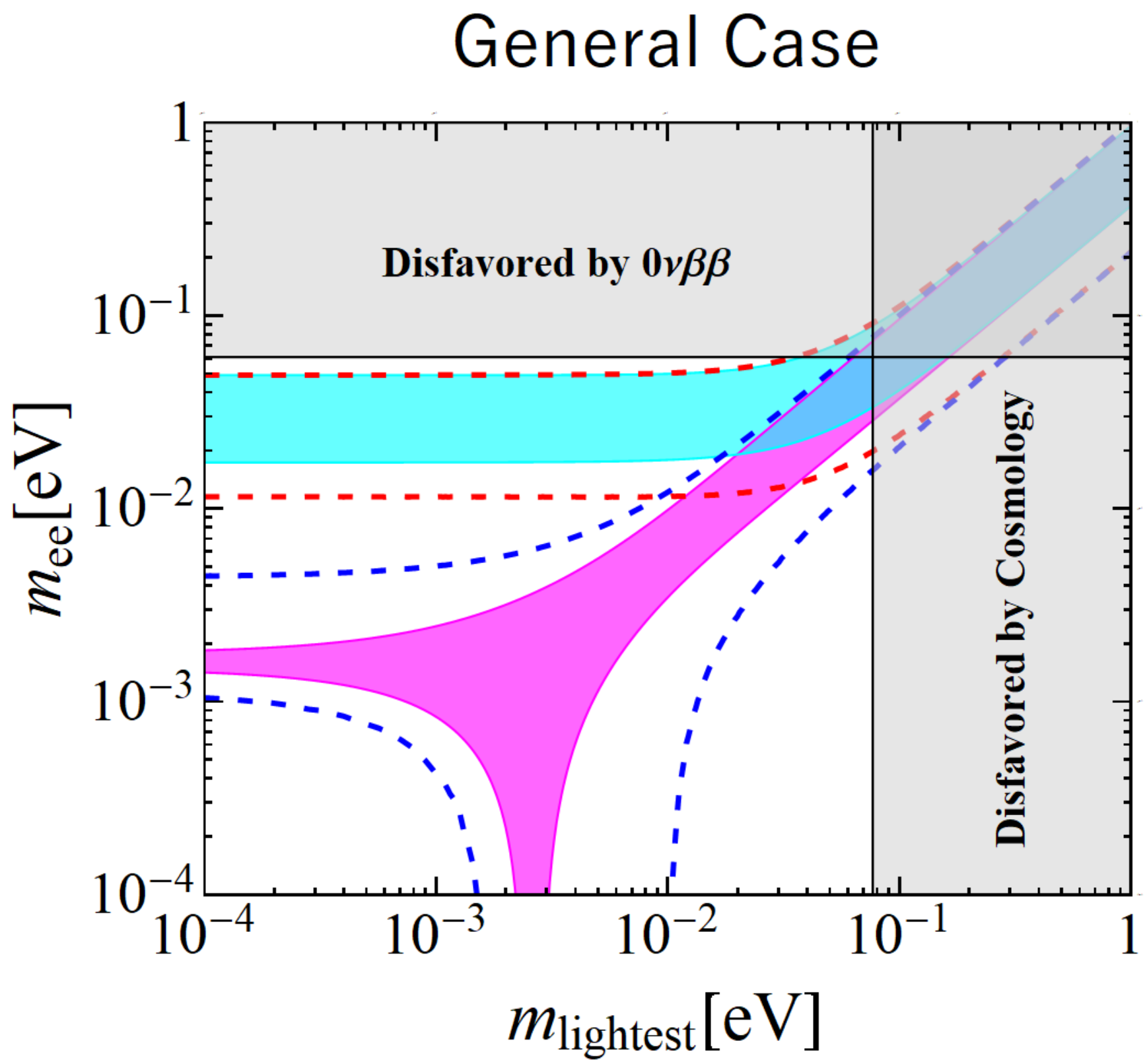}~~~~~~
\caption{$|m_{ee}|$ prediction for NO and IO in the most general gTBM ansatz. Here the parameters $\rho$ and $\theta$ are varied within their allowed $3\sigma$ ranges~\cite{deSalas:2017kay, globalfit}. Note that $m_{ee}$ does not depend on $\sigma$.}
\label{fig:mee-fig}
\end{figure}

From these mixing angles and phases in Eq.~\eqref{eq:mix23}, one can further obtain two non-trivial relations given by
\begin{eqnarray}
 \cos^2 \theta_{12} \, \cos^2 \theta_{13}  & = & \frac{2}{3} \, , \\
 \tan 2\theta_{23} \cos \delta_{CP} & = & \frac{5 \sin^2 \theta_{13} - 1}{4 \tan \theta_{12} \sin \theta_{13}}
\end{eqnarray}
The first is a correlation between $\theta_{12}$ and $\theta_{13}$, shown in Fig. \ref{fig:the12-the13}
while the second is a correlation between $\theta_{23}$ and $\delta_{CP}$, depicted in Fig. \ref{fig:the23-delta}.
Owing to the constrained nature of the mixing angles and phases of our ansatz, one also gets predictions for $m_{ee}$
shown in Fig. \ref{fig:mee-fig}.\\[-.2cm]

\begin{figure}[h]
 \centering
 \includegraphics[scale=0.5]{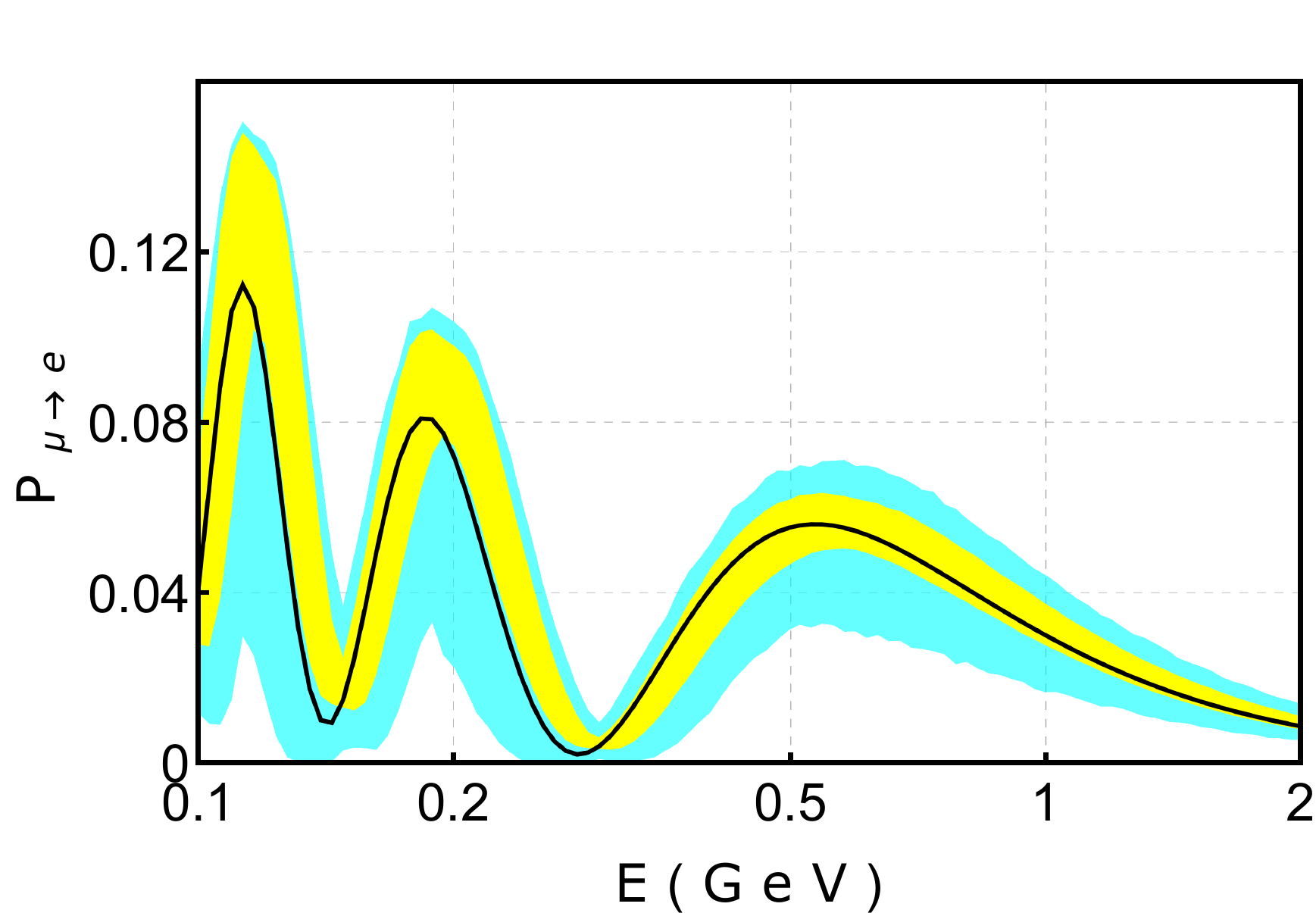}~~~~~~
\caption{The allowed range of electron neutrino appearance probability at T2K covers a more restricted region, thanks to the gTBM predictions.
Here black line corresponds to the best fit, the cyan region is the general three-neutrino result, while the yellow region is the gTBM prediction.}
\label{fig:osc-plot}
  \end{figure}

The predictions made by the gTBM ansatz can also be tested in currently running and upcoming neutrino oscillation experiments.
The predictions made by gTBM to oscillation experiments is illustrated in Fig.~\ref{fig:osc-plot}. This estimate is for the T2K setup, neglecting matter effects, as an approximation. Clearly the allowed range of electron neutrino appearance probability at T2K is substantially restricted w.r.t. the generic expectation.

In conclusion we have proposed a realistic generalization of the TBM ansatz which not only accounts for non-zero measured value of $\theta_{13}$ but also makes definite and testable predictions for the other parameters of the lepton mixing matrix, including CP phases.
Our gTBM matrix is characterized in terms of three independent parameters, which determine all six mixing parameters, leading to several testable predictions as we discussed at length.
Apart from correcting for $\theta_{13}$, the gTBM matrix retains many of the features of the original TBM matrix from point of basic underlying symmetries, as we showed by discussing various limits of the gTBM matrix.

Before closing we comment on the theoretical origin of the gTBM matrix. We note that this ansatz may be derived systematically by the method of generalized CP symmetries~\cite{Chen:2014wxa, Chen:2015nha,Yao:2016zev}. In this approach one starts from the TBM matrix and exploits various associated CP symmetries. 
For example, the mixing matrix in Eq.~\ref{eq:real-mu-tau} can be derived from the $S_4$ flavor symmetry and generalized CP~\cite{Feruglio:2012cw,Li:2013jya}.
A detailed derivation of the gTBM ansatz from the generalized CP approach, as well as other consequences of this methodology will be discussed elsewhere.

\begin{acknowledgments}

Work supported by the Spanish grants FPA2017-85216-P and SEV-2014-0398 (MINECO), PROMETEOII/2014/084 (Generalitat Valenciana), and by the National Natural Science Foundation of China, Grant No 11522546.

\end{acknowledgments}


%

\end{document}